\documentclass{tcibook}
\usepackage{fancyhea}
\usepackage{work}
\usepackage{bm}       
\usepackage{graphicx}
\usepackage{multirow}
\usepackage{lineno}
\usepackage{threeparttable}
\usepackage[normalem]{ulem}
\usepackage{color}
\usepackage[colorlinks = true,
            linkcolor = blue,
            urlcolor  = blue,
            citecolor = blue,
            anchorcolor = blue]{hyperref}

\def\beq{\begin{equation}}
\def\eeq#1{\label{#1}\end{equation}}
\def\eeqn{\end{equation}}

\newenvironment{Eqnarray}%
   {\arraycolsep 0.14em\begin{eqnarray}}{\end{eqnarray}}
\def\beqa{\begin{Eqnarray}}
\def\eeqa#1{\label{#1}\end{Eqnarray}}
\def\eeqan{\end{Eqnarray}}

\let\bar=\overbar

\def\lsim{\mathrel{\raise.3ex\hbox{$<$\kern-.75em\lower1ex\hbox{$\sim$}}}}
\def\gsim{\mathrel{\raise.3ex\hbox{$>$\kern-.75em\lower1ex\hbox{$\sim$}}}}

\def\del{\partial}
\def\Dslash{\not{\hbox{\kern-4pt $D$}}}
\def\dslash{\not{\hbox{\kern-2pt $\del$}}}
\def\pslash{\not{\hbox{\kern-2pt $p$}}}
\def\ETmiss{\not{\hbox{\kern-4pt $E$}}_T}

\def\Dlr{\mathrel{\raise1.5ex\hbox{$\leftrightarrow$\kern-1em\lower1.5ex\hbox{$D$}}}}

\def\MSB{{\bar{M \kern -2pt S}}}
\def\msb{{\bar{\scriptsize M \kern -1pt S}}}

\def\drb{{\bar{\scriptsize D \kern -1pt R}}}

\def\authorlist#1#2{
    \vskip 0.4in
\begin{center}\begin{large} {\bf  #1 } \end{large}
    \vskip 0.2in
              #2
     \vskip 0.2in
   \end{center}
}

\begin{document}


\pagenumbering{roman}

\parindent=0pt
\parskip=8pt
\setlength{\evensidemargin}{0pt}
\setlength{\oddsidemargin}{0pt}
\setlength{\marginparsep}{0.0in}
\setlength{\marginparwidth}{0.0in}
\marginparpush=0pt


\pagenumbering{arabic}

\renewcommand{\chapname}{chap:intro_}
\renewcommand{\chapterdir}{.}
\renewcommand{\arraystretch}{1.25}
\addtolength{\arraycolsep}{-3pt}
\setcounter{chapter}{4} 


\chapter{Report of the Topical Group on Micro-Pattern Gaseous Detectors for Snowmass 2021}

\authorlist{Conveners: B.~Surrow, M.~Titov, S.~Vahsen}{Community Contributors: A.~Bellerive, K.~Black, A.~Colaleo, K.~Dehmelt, K.~Gnanvo,\\  P.~Lewis, D.~Loomba, C.~O'Hare, M.~Posik, A.~White}

\begin{quote}
 This report summarizes white papers on micro-pattern gaseous detectors (MPGDs) that were submitted to the Instrumentation Frontier Topical Group IF05, as part of the Snowmass 2021 decadal survey of particle physics. An executive summary with key points is also provided.
 \end{quote}

\section{MPGDs: Executive Summary}

{\bf Background}

Gaseous Detectors are the primary choice for cost effective instrumentation of large areas and for continuous tracking of charged particles with minimal detector material. Traditional gaseous detectors such as the wire chamber, Resistive Plate Chamber (RPC), and time projection chamber (TPC) with multiwire proportional chamber (MWPC) readout remain critically important for muon detection, track-finding, and triggering in ongoing and planned major particle physics experiment, including all major LHC experiments (ALICE, ATLAS, CMS, LHCb) and DUNE.

Micro Pattern Gaseous Detectors (MPGDs) are gas avalanche devices with order $\mathcal{O}$(100~\textmu m) feature size, enabled by the advent of modern photolithographic techniques. Current MPGD technologies include the Gas Electron Multiplier (GEM), the Micro-Mesh Gaseous Structure (MicroMegas), THick GEMs (THGEMs), also referred to as Large Electron Multipliers (LEMs), the Resistive Plate WELL (RPWELL), the GEM-derived architecture (micro-RWELL), the Micro-Pixel Gas Chamber (\textmu-PIC), and the integrated pixel readout (InGrid).

MPGDs have already significantly improved the segmentation and rate capability of gaseous detectors, extending stable operation to significantly harsher radiation environments, improving spatial and timing performance, and even enabling entirely new detector configurations and use cases.

In recent years, there has therefore been a surge in the use of MPGDs in nuclear and particle physics. MPGDs are already in use for upgrades of the LHC experiments and are in development for future facilities (e.g., EIC, ILC, FCC, and FAIR). More generally, MPGDs are exceptionally broadly applicable in particle/hadron/heavy-ion/nuclear physics, charged particle tracking, photon detectors and calorimetry, neutron detection and beam diagnostics, neutrino physics, and dark matter detection, including operation at cryogenic temperatures. Beyond fundamental research, MPGDs are in use and considered for scientific, social, and industrial purposes; this includes the fields of material sciences, medical imaging, hadron therapy systems, and homeland security. 

{\bf Contributions to Snowmass}

Five commissioned white papers on MPGDs were developed during the 2021 Snowmass decadal survey. These summarize R\&D on MPGDs~\cite{Dehmelt:2022inw}, the future needs for MPGDs in nuclear physics~\cite{Barbosa:2022zql} and in three broad areas of particle physics: low-energy recoil imaging~\cite{OHare:2022jnx}, TPC readout for tracking at lepton colliders~\cite{Bellerive:2022wrb}, and tracking and muon detection at hadron colliders~\cite{Black:2022sqi}. A white paper with further details on a proposed TPC tracker for Belle~II was also submitted~\cite{Centeno:2022syq}.


{\bf Key Points}
The IF05 topical group would like to communicate the following high-level findings to the wider particle physics community:
\begin{itemize}
\item {\bf IF05-1}: Micro-pattern gaseous detectors (MGPDs) constitute an enabling technology that is key for large segments of the future U.S. NP and HEP programs, and which also benefits other communities. MPGDs provide a flexible go-to solution whenever particle detection with large area coverage, fine segmentation, and good timing is required.

\item {\bf IF05-2}: The technology is relatively young and should be advanced to performance limits to enable future HEP experiments. Support of generic and blue-sky R\&D is required to achieve this.
\item {\bf IF05-3}: The global HEP community would benefit from U.S. strategy coordination with the ECFA detector R\&D implementation process in Europe.
\item {\bf IF05-4}: In order to maintain and expand U.S. expertise on MPGDs, The U.S. NP and HEP communities would benefit strongly from a joint MPGD development and prototyping facility in the U.S.
\end{itemize}

\section{MPGDs: Recent advances and current R\&D}

 Recent developments in the field of MPGDs, and the role of the RD51 collaboration, are summarized in Ref.~\cite{Dehmelt:2022inw}. MPGDs were developed to cost-effectively cover large areas while offering excellent position and timing resolution, and the ability to operate at high incident particle rates. Significant development time was invested in optimizing manufacturing techniques for MPGDs, in understanding their operation, and in mitigating undesirable effects such as discharges and ion backflow. The early MPGD developments culminated in the formation of the RD51 collaboration hosted by CERN, which has become the critical organization for promotion of MPGDs and which coordinates all aspects of their production, characterization, simulation and use in an expanding array of experimental configurations. The CERN MPGD Workshop is a source
of essential expertise in production methods, mitigation and correction of manufacturing issues, and the development of MPGDs for specific experimental environments. 

An impressive array of MPGDs has been developed, from the initial GEM and Micromegas, now used in a wide variety of applications and configurations, through the more recent ThickGEMs, and microR-Wells with resistive layer(s) to mitigate discharge effects. MPGDs are now also used jointly with other detector elements, for example with optical readout and in liquid Argon detectors. In parallel with MPGD detector development, there has been an important creation of a standardized, general electronics system, the Scalable Readout System (SRS). This system has seen widespread use and is of great utility in allowing integration of a variety of frontend electronics into one data acquisition system. For Snowmass 2021, a number of Letters of Interest were received that illustrate ongoing developments and expansion of use of MPGDs. Here, we highlight high-precision timing, high-rate applications, expansion of the SRS readout system triggering capabilities, and reduction of ion backflow.

\begin{figure}[htp]
    \centering
    \includegraphics[trim={0 0.7cm 0 0},clip,width=0.9\textwidth]{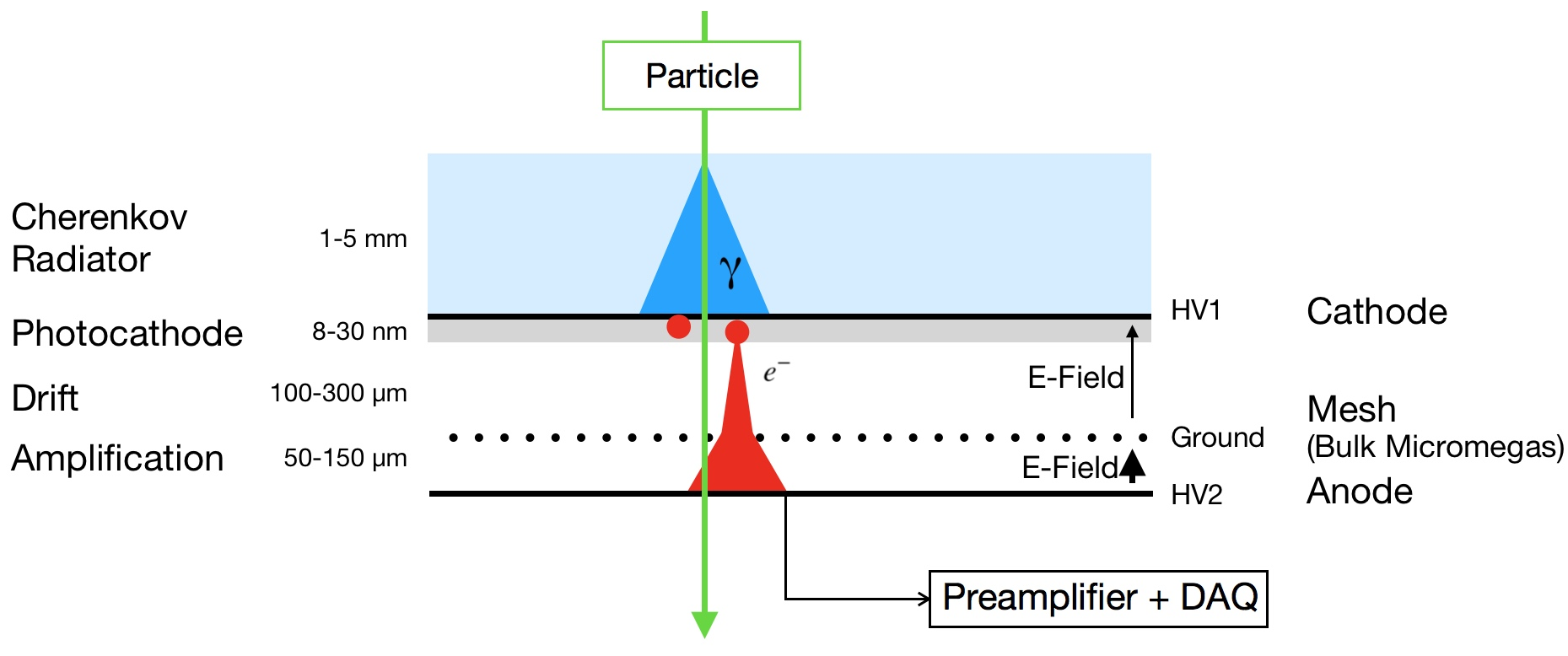}
    \caption{Sketch of the PICOSEC-Micromegas detector concept.~\cite{Dehmelt:2022inw}}
    \label{fig:picosec}
\end{figure}

{\bf Pico-second timing} The RD51 PICOSEC collaboration is developing very fast timing detectors using a two-stage system with a Cerenkov radiator producing photons to impact a photocathode. The photo-electrons are then drifted to and amplified in a Micromegas layer (Fig.~\ref{fig:picosec}). Prototypes have demonstrated single-photon timing at the 45~ps level and at the 15~ps level for MIPs. Multi-pad detectors are being developed, and studies applying an artificial neural network to the waveform have shown potential for very precise timing – in agreement with simulations. Potential applications of the PICOSEC technique include precise timing in electromagnetic showers, and time of flight systems for the future Electron Ion Collider.

{\bf High-rate applications} Micromegas detectors with a resistive layer for spark mitigation and small readout pads are in development for high rate operation. Several schemes for implementing the resistive layer have been tested. These include resistive pads covering the readout pads, layers of Diamond-like Carbon (DLC) across the plane of the detector, and hybrid schemes using both approaches. Prototype detectors have been tested using sources and X-rays in the RD51 Gaseous Detector Developme (GDD) laboratory, as well as particle (muon,pion) beams at CERN and PSI. Comparisons have been made of the rate capability of the various prototypes. Results are reported on the gain variations and operability at high rates of $\mathcal{O}$(10 MHz/cm$^2$) for pad and DLC prototypes. Energy resolution measured using X-rays is better for the DLC version than the pad version, attributable to the more uniform electric field in the former. Spatial resolution was measured using particle beams with the DLC configuration again performing better --- at the 100 micron level. Having demonstrated the desired rate capabilities and other characteristics of the small prototypes, a larger prototype is under construction, which also may include integration of the readout electronics with the detector.

{\bf Extended triggering capabilities}
There has been an evolution of the Scalable Readout System (from SRS to SRSe) to include realtime trigger functionality, deep trigger pipelines and a generalized frontend link using the eFEC, extended Frontend Concentrator backend. This will allow for the expanded use of a variety of frontend ASICs, and ability to use realtime triggering with firmware in an FPGA. A range of possible triggers is possible including, for example, hit combinations and energy sums. Ongoing plans for SRSe foresee script development for FPGAs, establishing a set of initial triggers, testing of the first eFECs, and addressing the needs of specific experiments.

{\bf Reduced ion backflow} New structures are being designed to restrict ion backflow, to limit performance degradation and/or detector damage. Specifically, multi-mesh MPGD structures, which preserve the desirable features of Micromegas while controlling ion backflow, have been studied. Results with double and triple mesh structures, with adjusted and optimized gaps, have shown that high gain and very low ion backflow can be achieved offering the prospect of excellent performance in future detector applications.

Many other improvements of MPGDs are being actively pursued. Several of these efforts are commented on below, in the context of their intended applications.

\section{MPGDs for nuclear physics experiments}
Many current and future nuclear physics (NP) experiments in the United States have or are implementing MPGDs for tracking and particle identification (PID) purposes. Here, we summarized the role that MPGDs play in NP experiments, and the R\&D needed to meet the requirements of future NP experiments. More detail can be found in Ref.~\cite{Barbosa:2022zql}. 

{\bf Advanced MPGDs for Tracking at the Electron-Ion Collider}
The physics program of the Electron-Ion Collider (EIC), to be built at Brookhaven National Laboratory, requires its tracking system to have low mass ($ X/X_0 ?1\% $), large area $\mathcal{O}$(1~m$^2$), and excellent spatial resolution $\mathcal{O}$(100~\textmu m). MPGDs such as the GEM, Micromegas, and $\mu$RWELL can meet these requirements. Furthermore, the EIC is expected to have relatively low rates, below 100~kHz/cm$^2$, which is well within the operating range of current MPGDs. Current R\&D focuses on reducing large-area detector material budgets and the number of channels needed to be read out while maintaining excellent spatial resolution. The EIC can benefit from future R\&D, which aims to reduce the detector's material and service budgets further, to achieve a spatial resolution of order $\sim 20$~\textmu m, and to implement particle identification capabilities into MPGD detectors. 

{\bf MPGD Technologies for Particle Identification in Nuclear Physics Experiments}
PID plays a vital role in high energy physics (HEP) and NP physics. The next generation of high-intensity accelerators and increased demand for precision measurements will require the development of high granularity detectors, such as those based on MPGD technologies. Combining a high-precision tracker with PID-capable technology could prove valuable for future experiments. 
A high-precision MPGD tracker combined with a transition radiation (TR) option for particle identification could provide important information necessary for electron identification and hadron suppression. A radiator installed in front of an MPGD entrance window provides an efficient yield of TR photons. 
MPGD-based photon detectors offer the ability to provide PID through Cerenkov imaging techniques. Such detectors are attractive for experiments like the EIC as they can provide a cost-efficient option for large-area, low-material budget detectors that can operate in a magnetic field.

{\bf MPGD Technologies for Low Energy Nuclear Physics at FRIB}
The Facility for Rare Isotope Beams (FRIB) at Michigan State University (MSU) will become the world's most advanced facility for the production of rare isotope beams (RIBs). With the delivery of beams starting in Spring 2022, FRIB will be capable of producing a majority (around 80\%) of the isotopes predicted to exist, including more than 3,000 new isotopes, opening exciting perspectives for exploring the uncharted regions of the nuclear landscape. Its scientific impact will span a better understanding of open quantum systems at the limits of stability through investigations of the structure and reactions of atomic nuclei and their roles in nuclear astrophysics, low-energy tests of fundamental symmetries, and practical applications that benefit humanity. MPGD technologies play an essential role in the science program's success at FRIB. Applications of MPGD technologies include low-pressure tracking and particle-identification (PID) at the focal planes of magnetic spectrometers, Active-Target Time-Projection-Chambers (TPCs), and TPCs for the detection of exotic decay modes with stopped RIBs. The unprecedented discovery potential of FRIB can be achieved by implementing state-of-the-art experimental equipment and overcoming the challenges of current devices by taking the following measures: improving spatial and energy resolutions, optimizing pure-gas operation for active target mode, improving reliability and radiation hardness at a lower cost, reducing ion-back flow to minimize secondary effects and increase counting rate capability, and integrating electronic readout to reach high channel density, fast data processing, and storage.

{\bf MPGD Technologies for Nuclear Physics at Jefferson Lab}
Future spectrometers for NP experiments at the Thomas Jefferson National Accelerator Facility (Jefferson Lab) require large area $\mathcal{O}$(m$^2$),  low mass (X/X$_0 \leq$ 1\%), excellent spatial resolution $\mathcal{O}(100\, $\textmu m), excellent timing $\mathcal{O}$(10 ns), high rate $\mathcal{O}$(1~MHz/cm$^2$) tracking detectors technologies for operation in high background rate and high radiation environment. Only MPGD technologies such as GEMs, Micromegas, or $\mu$RWELL detectors can satisfy the challenges of high performances for large acceptance at reasonably low cost. Critical R\&D for the next decades will focus on new ideas to develop ultra-low mass, large area, and radiation tolerant MPGD trackers with even higher rate capabilities. The performance of new materials (Chromium GEMs, Aluminum-based readout strips) and original concepts for anode readout such as capacitive, resistive, and zigzag readouts for high-performance \& low channel count MPGD detectors will be explored.

\begin{figure}[htp]
    \centering
    \includegraphics[trim={0 0.7cm 0 0},clip,width=0.9\textwidth]{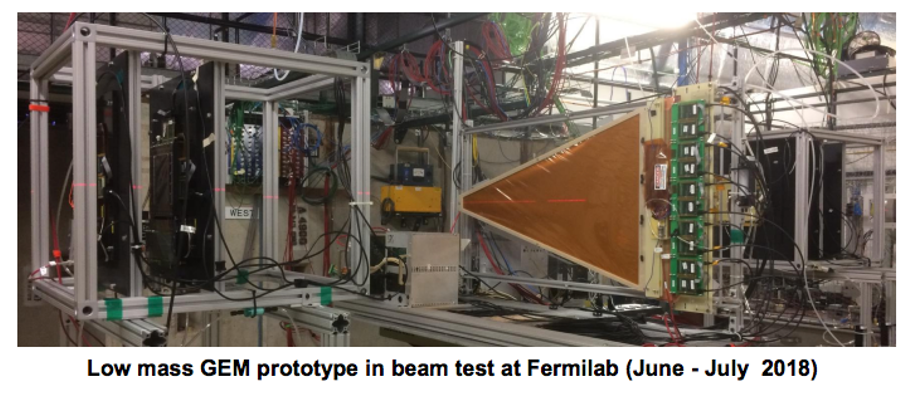}
    \caption{Low mass GEM prototype in beam test at Fermilab, June -- July 2018~\cite{Barbosa:2022zql}}
    \label{fig:testbeam}
\end{figure}

\paragraph{Electronics, DAQ, and Readout Systems for MPGD Technologies}
The EIC will implement a full streaming readout architecture. This trigger-less implementation will consist of front-end circuitry and processors to enable data collection, processing, and analysis: front-end ASICs will be designed to meet wide bandwidth sub-detector and system requirements; front-end processors will include FPGAs to provide data aggregation and enable flexible algorithms to reduce data volume while maintaining wide system bandwidth; system clock distribution with timing precision of the order of 1 ps; link exchange modules and servers for data processing; and data transport via extensive use of optical fibers. The development of ML/AI algorithms will play a critical role in enabling a full detector bandwidth of 100 Tbps and delivering data output rates of 100 Gbps.

The readout of MPGD detectors requires specific front-end ASICs to amplify and digitize the detector signals with performance requirements depending on their constraints and application. The ASICs should also be compatible with the high-speed streaming readout DAQ systems considered for future experiments at EIC and elsewhere. Existing chips (like SAMPA and VMM) partially satisfy these requirements and can be used for specific applications. Nevertheless, an initiative has been launched to develop a new versatile SALSA ASIC that satisfies the requirements of most MPGD applications in HEP, including both streaming and triggered readout paradigms.

{\bf A dedicated MPGD Development Facility} The U.S. NP efforts listed above would  benefit greatly from a U.S. based MPGD development facility, as highlighted in the executive summary.

\section{MPGDs for recoil imaging in dark matter and neutrino experiments}

\begin{figure}
\begin{center}
\includegraphics[height=0.35\columnwidth]{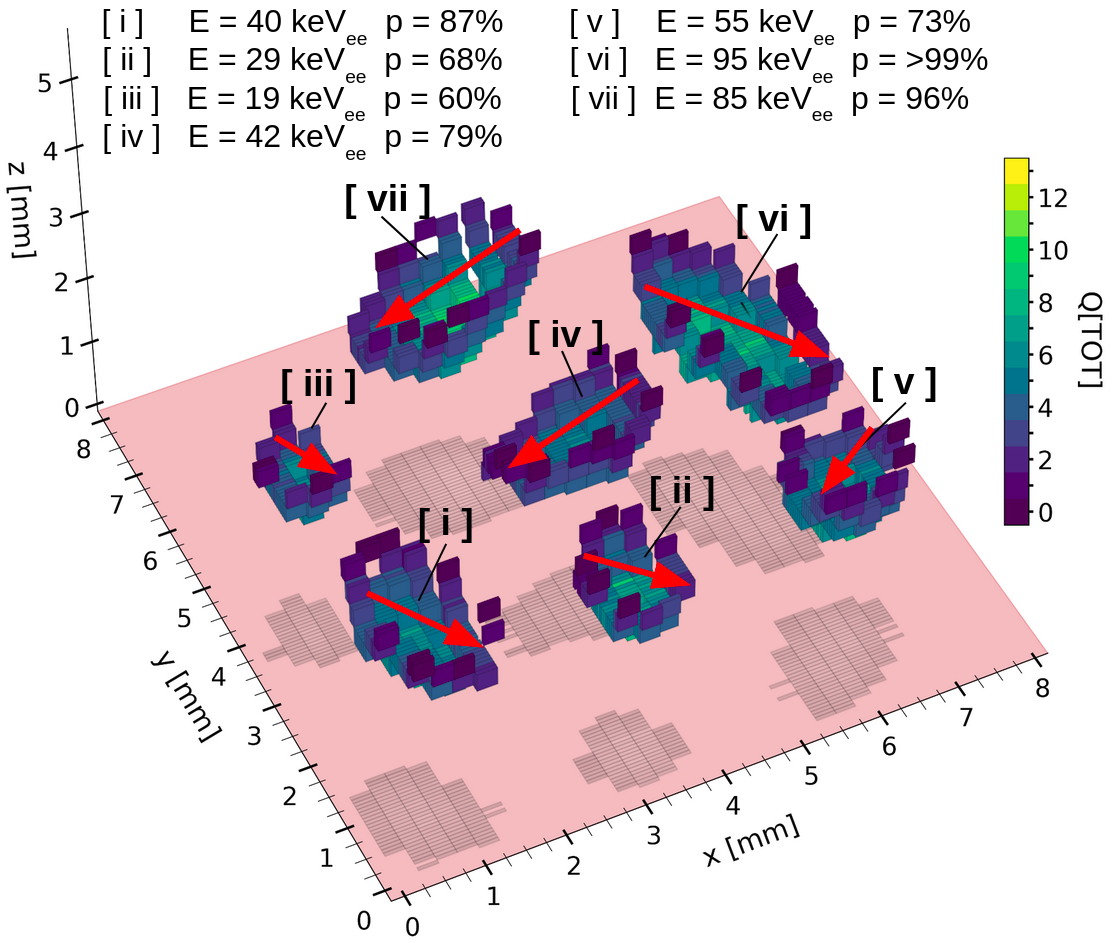}
\includegraphics[height=0.35\columnwidth]{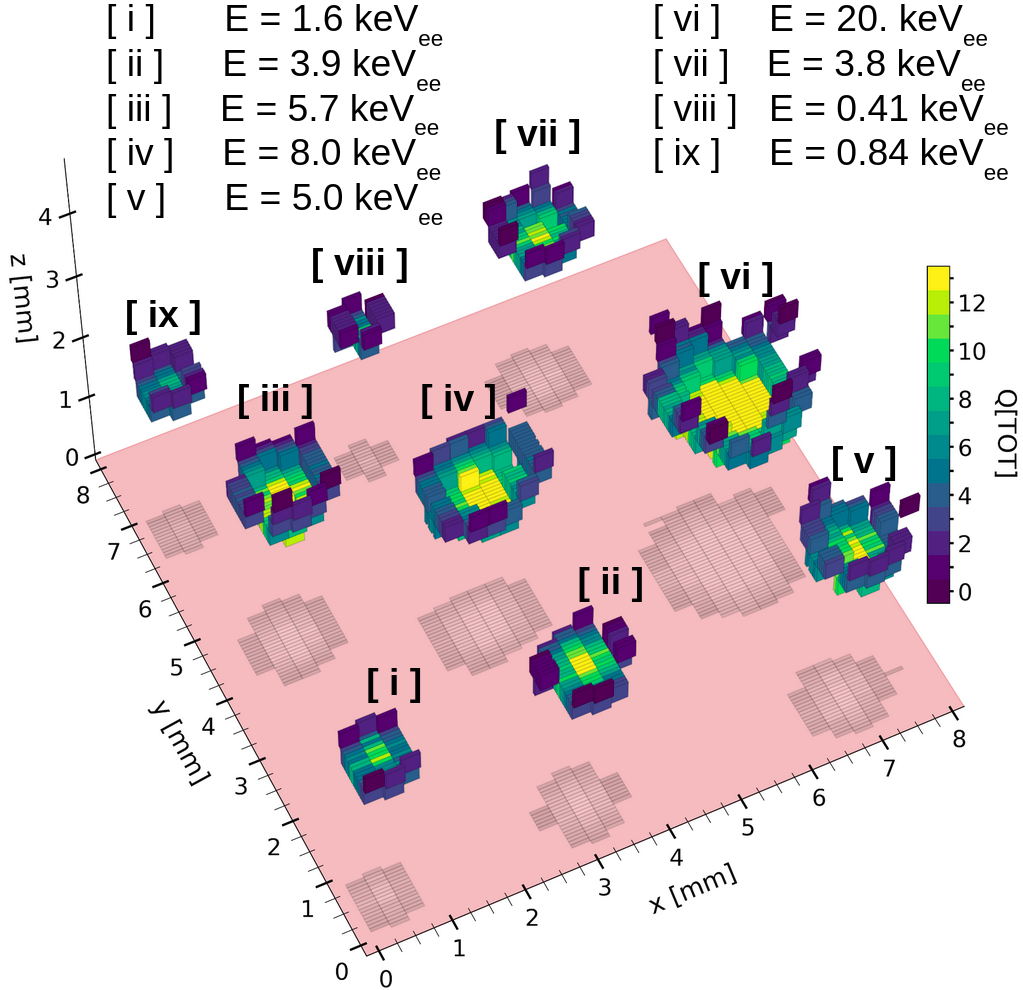}
\caption{Helium recoils measured in a gas TPC with GEM amplification and pixel ASIC charge readout, in 760 Torr He:CO$_2$ (70:30) gas. Left: 3d directional reconstruction with the detector in low-gain mode, down to 20~keV$_{\rm ee}$. The red arrows show fitted recoil directions, with the head and tail (i.e. sign of the vectors) determined by a 3D convolutional neural network. Right: Demonstration of detection down to sub-keV energies, with the detector in high-gain mode.}\label{gas_tpc_events}
\end{center}
\end{figure}

MPGDs can be used to read out the ionization in low-density gas time projection chambers (TPCs) with exquisite sensitivity and spatial resolution. In the most advanced MPGD TPCs, even individual primary electrons---corresponding to an energy deposit on the order of $\sim$30~eV---can be detected with negligible background from noise hits, and 3D ionization density can be imaged with $\sim$(100~\textmu m)$^3$ voxel size, as shown in Fig.~\ref{gas_tpc_events}. This new experimental technique, enabled by MPGDs, has a large number of interesting applications in fundamental and applied physics. We briefly discuss examples below. Further detail on this emerging field can be found in Ref.~\cite{OHare:2022jnx}.

$\bf CYGNUS$ Early R\&D has established high-definition gas TPCs (HD TPCs) with MPGD readout as the leading candidate technology for imaging the short, mm-scale tracks resulting from keV-scale nuclear recoils in gas. In this context, the detailed ionization images can distinguish electronic from nuclear recoils with high confidence, and can provide the 3D vector direction (i.e.~both the recoil axis and the head/tail assignment) of both types of recoils, even at the 10-keV-scale recoil energies relevant to dark matter (DM), neutrino-nucleus scattering, and more. 

One of the most intriguing applications is to scale up TPCs with MPGD readout to construct a competitive, low-background, high-definition ionization imaging experiment. By virtue of a planned ultimate sensitivity approaching a single primary electron, the proposed CYGNUS experiment would be sensitive to any process---whether known, hypothesized, or not yet thought of---that produces ionization in the target gas. 

While much of R\&D on MPGDs occurs in the context of the RD51 collaboration at CERN, several key advances in the applications of MPGDs to low-energy physics were made in the US. There is an opportunity here for the US to take the lead and initiate a novel experimental program focused on recoil imaging, with broad scientific scope, one that we have only just started to map out. Because large areas, $\mathcal{O}$(1000~m$^2$), of MPGDs are required, there are clear synergies with this proposal and the Electron Ion Collider, where MPGDs will be used broadly, and where R\&D, production, and test facilities in the US are also desireable. 

{\bf Required R\&D} To enable and optimize the sensitivity of a large-scale recoil imaging experiment like CYGNUS, several major R\&D goals should be pursued in the next decade. First, HD TPCs should be advanced to their natural performance limit, where primary electrons are {\it counted individually} in 3D with $\mathcal{O}$(100~\textmu m)$^3$ spatial resolution. In this regime, the energy resolution is expected to reach a fundamental limit, finite dynamic range of detectors is mitigated, and particle identification and directional capabilities required for physics measurements will be maximized.

The second step is to enable this level of performance in larger detectors, at reasonable cost. Electronic readout in the form of Micromegas detectors with 2D $x$/$y$ strips are a candidate technology for this, and the main approach being pursued in the US. Key ingredients to this include self-triggering, highly-multiplexed electronics with topological programmable triggers. Another main direction is optical readout, which is the main approach pursued in Europe. In place of electron drift, \emph{negative ion drift} (NID) will also likely be required to reduce diffusion and enable 3D fiducialization. Depending on how negative ions are used in detail, custom front-end electronics may then be required. It is desirable to achieve NID with gases that have low environmental footprint, and capabilities to clean and recirculate gas are already being developed. 

The level of internal radioactivity in relevant MPGD technologies must be reduced for an experiment with large exposure. The exact level required also depends strongly on the particle identification capabilities of the detector. Early R\&D has shown that modern machine learning can make a large difference in this context. For example, 3D convolutional neural networks are ideally suited to analyze the 3D ionization images created by HD TPCs. This can improve performance by up to three orders of magnitude, thereby lowering the requirement on radiopurity by the same factor. Algorithm development and machine learning both for offline analysis and for smart triggers (known as ``machine learning on the edge'') are therefore a crucial part of our proposed program.

{\bf IAXO} A notable application of recoil imaging in MPGDs beyond DM and neutrinos is for the International Axion Observatory (IAXO). IAXO will be an axion helioscope, an experiment that aims to detect the keV-scale photons generated when the hypothesized flux of axions coming from the Sun enters a large static magnetic field. IAXO is the proposed successor to the CAST experiment, and its intermediate stage BabyIAXO plans to begin data-taking in 2025--2026. A range of technology is planned to be tested for IAXO to further improve backgrounds levels and discrimination capabilities, including novel devices such as GridPix and Timepix3.

{\bf Other applications} Recoil imaging is also a desirable strategy for background rejection and signal identification in applications totally apart from those listed already. Directional neutron detection, the measurement of the Migdal effect, X-ray polarimetry, and the detection of rare nuclear decays, are to name just a few.

{\bf High density gases and dual readout} One final concept that has attracted some interest recently, is the use of high density gases such as SeF$_6$ or argon. Several groups are exploring TPC designs that can provide the necessary sub-mm to 10~micron resolution with such gases. One potential way this could be achieved is via the use of a `dual readout' TPC which can detect both the positive ions as well as the electrons generated by a recoil event. TPCs using gaseous argon could be of interest for studies of the neutrino sector, for example $\tau$-tracking for the study of $\nu_\tau \tau$ charged current interactions.

\section{MPGDs at future high energy physics colliders}

Advances in our knowledge of the structure of matter during the past century were enabled by the successive generations of high energy particle accelerators, as well as by continued improvement in detector technologies. In this context, MPGDs have become a preferred solution for enabling both continuous, low-mass charge-particle tracking in TPCs and large-area muon detection systems. These topics are covered in two dedicated Snowmass Whitepapers~\cite{Bellerive:2022wrb,Black:2022sqi}, summarized below. A dedicated Snowmass Whitepaper focused on a proposed TPC tracker for Belle~II is also available~\cite{Centeno:2022syq}.

{\bf TPCs} The physics goals of a future Higgs factory and also
at the flavour-precision frontier, have put stringent constraints on the need to develop novel instrumentation. Time Projection Chambers (TPCs) operating at $e^+e^-$ machines 
in the 1990's reached their sensitivity limit and new
approaches needed to be developed to overcome the need for improved resolution. The spatial and timing resolution goals nowadays represent an order of magnitude improvement 
over the conventional proportional wire/cathode pad TPC performance, which is limited by the intrinsic $\bf{E} \times \bf{B}$ effect near the wires, and approaches the fundamental limit imposed by diffusion. 
Other detrimental effects such as material budget, cost per readout channel and power consumption also represent serious challenges for future high-precision tracking detectors. 
One of the most promising areas of R\&D in subatomic physics is the novel development of gaseous detectors. Micro Pattern Gas Detector (MPGD) technologies have become a well-established 
advancement in the deployment of gaseous detectors because those will always remain the primary choice whenever large-area coverage with low material budget is required for particle detection.  
MPGDs have indeed a small material budget,  which is important in a high background or a high-multiplicity environment, and naturally reduce space charge build up in the drift volume by 
suppressing positive ion feedback from the amplification region. Of greatest importance however, is that the $\bf{E} \times \bf{B}$ effect is negligible for an MPGD because the micro holes 
have $\sim$100 $\mu$m spacing, which offers a rotationally symmetric distribution and thus no preferred track angle. 

Many detector designs aimed at future lepton colliders utilize MPGD devices. We focus on three proposed MPGD-based TPCs: (i) the tracker of the International Large Detector (ILD) at the International Linear Collider (ILC), (ii) a potential replacement of the wire chamber of the Belle II detector at the SuperKEK B-Factory with a TPC, and (iii) a TPC for a detector at the Circular Electron Positron Collider (CEPC). 

Overall, an MPGD-based TPC offers excellent tracking performance, while enabling continuous or power-cycled readouts. Historically, TPCs were the main central tracking 
chambers of ALEPH and DELPHI at the electron-positron collider LEP, where Americans were collaborators. The T2K  Near Detector with Micromegas represents another 
area where TPC technology was deployed with engagement of participants from North America. The upgrade of the ALICE TPC is a more recent example of the usage of MPGDs 
with participation from institutions from the United States. The ALICE main central-barrel tracking used to rely on multi-wire proportional chambers, 
which have since been replaced by a TPC with GEM readouts designed in an optimized multilayer configuration, which stand  up to the technological challenges imposed by continuous TPC operation at high rates. The requirement to keep the ion-induced space-charge distortions at a tolerable level, which leads to an upper limit of 2\% for the fractional ion backflow, has been achieved. The
upgraded TPC readout will allow ALICE to record the information of all tracks produced in 
lead-lead collisions at rates of 50 kHz, while producing data at a staggering rate of 3.5 TB/s. For both the T2K and ALICE TPCs, partnership with CERN allows the fabrication of 
anode boards of size of order of 50 cm x 50 cm. 

The TPC concept is viewed in particle physics as the ultimate drift chamber since it provides 3D precision tracking with low material budget and enables particle identification through dE/dx 
measurements with cluster counting techniques. At ILC and CEPC, as well as for Belle II upgrades, MPGD TPC technologies are the preferred main tracking system for some conceptual
detectors. There are synergies with other MPGD detector activities (as summarized here) that offer
clear motivation for gaseous tracking at lepton colliders. Gaseous tracking devices have been extremely successful in providing precision 
pattern recognition. They provide hundreds of measurements on a single track, with an extremely low material budget in the central region of the detector. This results in accurate 
track reconstruction and hence high tracking efficiency. The continuous measurements of charged particle tracks allow for precise particle identification capabilities, which have the possibility not
only to achieve excellent continuous tracking, but also to improve jet energy resolution and flavour-tagging capability for an experiment at a future lepton collider. These are two essential 
advantages for experiments at a lepton collider. The main challenges for the design of a 
large TPC are related to the relatively high magnetic field in which planned detectors will operate. For accurate measurements of the 
momenta of charged particles, the electromagnetic field has to be known with high precision. Final and sufficient calibration of the field map can be achieved using corrections derived 
from the events themselves, or from dedicated point-like and line sources of photoelectrons produced by targets
located on the end-plates when illuminated by laser systems. While the event rate at lepton collider detectors can easily be accommodated by current TPC readout technology, R\&D to mitigate the effects of secondary processes from bunch-bunch interactions is ongoing. MPGD technologies offer a wide-range of applications and call for synergy in detector R\&D at future 
lepton colliders. The availability of a highly integrated 
amplification system with readout electronics allows for the design of gas-detector systems with channel densities comparable to that of modern silicon detectors. 
This synergy with silicon detector ASIC development is very appealing for MPGD TPCs since recent wafer post-processing enables the integration of gas-amplification 
structures directly on top of a pixelized readout chip. 

The {\bf ILD TPC, LCTPC}, is based on mature hardware and software contributions from multiple partners
and in particular from the United States ($e.g.$ Cornell University and Wilson Laboratory - now the Cornell Laboratory for  Accelerator-Based Sciences and Education).
LCTPC is conceptually ready as it meets performance and engineering requirements. It is the outcome of decades of research and innovation in MPGDs.
Single-hit transverse resolution results from testbeam at 1 T magnetic field extrapolated
to the 3.5~T field of ILD clearly demonstrate that single point resolution of 100~\textmu m after $\sim$2 m of drift over about 200 measurement points 
is achievable with several MPGD technologies (GEM, Micromegas or GridPix). LCTPC achieved unprecedented spatial resolution of 35~\textmu m at zero drift distance for 2~mm wide readout pads, a world record, and 55~\textmu m with 3~mm wider pads in a high field magnet. This translates to two-hit separation of $\sim$2~mm and a momentum resolution of $\delta(1/p_T) \simeq 10^{-4} /$ GeV/c (at 3.5 T), which are 
the required performance of the TPC as a standalone tracker at ILD for ILC. Other areas of MPGD developments are ongoing on ion gating, dE/dx, power-pulsed electronics and cooling. 

Similar simulations were performed by members of the {\bf Belle II} Collaboration showing 
that a GridPix-based TPC could be the ultimate central tracker for an upgrade detector at a future ultra-high luminosity B-Factory. The readout choice will need to be adapted to the
beam structure of an ultra-high luminosity SuperKEKB upgrade, and probably a buffer that can handle discrete readout of multiple concurrent events will be required.
The baseline design of a CEPC detector is an ILD-like concept, with a 
superconducting solenoid of 3.0 Tesla (Higgs run) and 2.0 Tesla (Z pole run) surrounding the inner silicon detector, the TPC tracking
detector and the calorimetry system.
The {\bf CEPC TPC} detector will operate in continuous mode on
the circular machine. As for the ILD TPC, MPGD technologies are applicable and desirable for a detector at CEPC.

{\bf MPGDs for large-scale muon detectors at colliders} Gaseous detectors are the primary choice for cost effective instrumentation of very large areas, with high detection efficiency in a high background and hostile radiation environment, needed for muon triggering and tracking at future facilities.
They can provide a precise standalone momentum measurement or be combined with inner detector tracks resulting in even greater precision. Adding precise timing information ,$\mathcal{O}$(ns), allows control of uncorrelated background, mitigates pile-up and allows detection of extremely long lived particles that behave like slow muons propagating through the  detector volume over a time as long as a few bunch crossings.

 With the invention and evolution of MPGDs during the last twenty years, gaseous detectors improved significantly in spatial resolution and rate capability. 
MPGDs allow stable operation at very high background particle flux with high efficiency and excellent spatial resolution. These features determine the main applications of these detectors in particle physics experiments as precise muon tracking in high radiation environment as well as muon tagger and trigger in general purpose detectors at HEP colliders. Two of the most prominent MPGD technologies, the GEM and MicroMegas, have been successfully operated in many different experiments, such as Compass, LHCb, and TOTEM.
In addition, the low material budget and the flexibility of the base material makes MPGDs suitable for the development of very light, full cylindrical fine tracking inner trackers at lepton colliders such as KLOE-2 at DAFNE (Frascati, IT) and BESIII at BEPCII (Beijing, CN).

A big step in the direction of large-size applications has been obtained both with conceptual consolidation and industrial and cost-effective manufacturing of MPGDs by developing new fabrication techniques: resistive Micromegas (to suppress destructive sparks in hadron environments) and single-mask and self-stretching GEM techniques (to enable  production  of  large-size  foils  and  significantly reduce detector  assembly time). Scaling up of MPGDs to very large single unit detectors of $\mathcal{O}$(m$^2$), has facilitated their use in muon systems in the LHC upgrades. 
Major developments in the MPGD technology have been introduced for the ATLAS and CMS muon system upgrades, towards establishing technology goals, and addressing engineering and integration challenges. Micromegas and GEM have been recently installed  in the ATLAS New Small Wheel, CMS GE1/1 station respectively, for operation from Run 3 onward, as precise tracking systems. Those radiation hard detectors, able to cope with the expected increased particle rates, exhibit good spatial resolution, $\mathcal{O}$(100 \textmu m) and have a time resolution of 5--10~ns. 
In the CMS muon system additional stations, GE2/1 and ME0, based on GEMs with high granularity and spatial segmentation, will be installed to ensure efficient matching of muon stubs to offline pixel tracks at large pseudo-rapidities during HL-LHC operation. 
Several solutions 
($\mu$-RWELL, Micro Pixel Chamber ($\mu$-PIC), and small-pad resistive Micromegas were also considered for the very forward muon tagger in the ATLAS Phase-II Upgrade Muon TDR proposal. Here, the main challenges are discharge protection and miniaturization of readout elements, which can profit from the ongoing developments on Diamond-Like Carbon (DLC) technology. The $\mu$-RWELL is the baseline option for the Phase-II Upgrade of the innermost regions of the Muon System of the LHCb experiment (beyond LHC LS4).

The new era of Particle Physics experiments is moving towards the upgrade of present accelerators and the design of new facilities operating at extremely high intensities and particle energies such as the \textbf{Future Circular Colliders} and the \textbf{Muon Collider}.
Cost effective, high efficiency particle detection in a high background and high radiation environment is fundamental to accomplish their physics program.
Different critical aspects such as the high particle rates, discharge probabilities and accumulated doses expected at future colliders must be taken into account. Modifications or new detector configurations are to be investigated by relying  on innovative technological solutions.
Muon systems at future lepton colliders, (ILC, CLIC, CepC, FCC-ee, SCTF) or \textbf{LHeC}, do not pose significant challenges in terms of particle fluxes and the radiation environment. Therefore many existing MPGD technologies are suitable for building future large muon detection systems. For example the $\mu$RWELL technology is envisaged to be utilized for the muon detection system and the preshower detector of the IDEA detector concept that is proposed for the FCC-ee and CepC future large circular leptonic colliders. In addition $\mu$RWELL are candidates for the inner tracking system at future high luminosity tau-charm factories, STCF in Russia and SCTF in China.
Generally, background rates in LHeC muon detector, which are based on the updated design of ATLAS Phase-II Muon spectrometer, are lower than in $pp$ colliders. 
On the other hand, the expected particle rates for the muon tracking and triggering at future \textbf{hadron colliders, such as the FCC-hh}, make the existing technologies adequate in most regions of the spectrometers, but require a major R\&D  for the very forward endcap region.
In a \textbf{multi-TeV muon collider}, the effect of the background induced by the muon beam decays is extremely important, since it can contaminate the Interaction Region (IR) from a distance that varies with the beam energy, the collider optics and the superconducting magnets. Therefore, the rate of background is particularly relevant in the forward region. Tracking and triggering can be obtained with multi-layer structures, for an efficient local muon segment reconstruction. A new generation Fast Timing MPGD (FTM, Picosec) is considered to mitigate the beam induced background, by rejecting hits uncorrelated in time.

MPGDs offer a diversity of technologies that allow them to meet the required performance challenges at future facilities and in various applications, thanks to the specific advantages that each technology provides. On-going R\&D should focus on pushing the detector performance to the limits of  each technology by overcoming the related technological challenges. 

\textbf{Required R\&D} should focus on stable operation of large area coverage, including precision timing information to ensure the correct track-event association, and on the ability to cope with large particle fluxes, while guaranteeing detector longevity using environmentally friendly gas mixtures and optimized gas consumption (gas recirculating and recuperation system). 
Strong constraints on response stability, discharge probability and space charge accumulation require  innovative technological solutions and novel detector configurations. Considering the high rate exposure of the detectors and the radiation hazards at future colliders, very strong restrictions to access the detector for reparations and replacement are expected. In this scenario long term operation requirements have to be guarantee also in term of mechanical and electronics robustness. The main challenges include gas tightness, over-pressure operation and electronics cooling.
Integration aspects have also to be optimised for easy accessibility and replaceability in complex installations. The assembly of a large scale detector components will require engineering effort to ensure mechanical precision. 
MPGDs require dedicated front-end electronics (FEE) development, both discrete and integrated (ASIC), focused on specific applications, while meeting a large set of challenging requirements such as: fast timing, large input capacitance, low noise, input discharge protection, cross-talk reduction, pixel size, compactness, low power consumption and detector integration.





\bibliographystyle{JHEP}
\bibliography{mybibliography} 

\providecommand{\href}[2]{#2}\begingroup\raggedright\begin{thebibliography}{1}

\bibitem{Dehmelt:2022inw}
K.~Dehmelt et~al., \emph{{Snowmass 2021 White Paper Instrumentation Frontier 05
  - White Paper 1: MPGDs: Recent advances and current R\&D}},
  \href{https://arxiv.org/abs/2203.06562}{{\ttfamily 2203.06562}}.

\bibitem{Barbosa:2022zql}
F.~Barbosa et~al., \emph{{Snowmass 2021 Instrumentation Frontier (IF5 - MPGDs)
  -- White Paper 2: Micro Pattern Gaseous Detectors for Nuclear Physics}},
  \href{https://arxiv.org/abs/2203.06309}{{\ttfamily 2203.06309}}.

\bibitem{OHare:2022jnx}
C.A.J.~O'Hare et~al., \emph{{Recoil imaging for dark matter, neutrinos, and
  physics beyond the Standard Model}},  in \emph{{2022 Snowmass Summer Study}},
  3, 2022 [\href{https://arxiv.org/abs/2203.05914}{{\ttfamily 2203.05914}}].

\bibitem{Bellerive:2022wrb}
A.~Bellerive et~al., \emph{{MPGDs for TPCs at future lepton colliders}},
  \href{https://arxiv.org/abs/2203.06267}{{\ttfamily 2203.06267}}.

\bibitem{Black:2022sqi}
K.~Black et~al., \emph{{MPGDs for tracking and Muon detection at future high
  energy physics colliders}},  in \emph{{2022 Snowmass Summer Study}}, 3, 2022
  [\href{https://arxiv.org/abs/2203.06525}{{\ttfamily 2203.06525}}].

\bibitem{Centeno:2022syq}
A.L.~Centeno, C.~Wessel, P.M.~Lewis, O.~Hartbrich, J.~Kaminski, C.~Mari\~nas
  et~al., \emph{{A TPC-based tracking system for a future Belle II upgrade}},
  in \emph{{2022 Snowmass Summer Study}}, 3, 2022
  [\href{https://arxiv.org/abs/2203.07287}{{\ttfamily 2203.07287}}].

\bibitem{Detector:2784893}
E.D.R.R.P.~Group, \emph{{The 2021 ECFA detector research and development
  roadmap}},  Tech. Rep. }, Geneva (2020),
  \href{https://doi.org/10.17181/CERN.XDPL.W2EX}{DOI}.

\end{thebibliography}\endgroup



\providecommand{\href}[2]{#2}\begingroup\raggedright\endgroup



\providecommand{\href}[2]{#2}\begingroup\raggedright\endgroup



\providecommand{\href}[2]{#2}\begingroup\raggedright\endgroup




\end{document}